\documentclass[aps,prb,showpacs,twocolumn,floats]{revtex4}
\usepackage{amssymb}
\usepackage{amsbsy}
\usepackage{amsmath}
\usepackage{epsfig}

\begin{document}

\title {Phases and collective modes of hardcore Bose-Fermi mixture in an optical lattice}

\author{S. Sinha$^{(1)}$ and K. Sengupta $^{(2,3)}$}

\affiliation{$^{(1)}$ Indian Institute of Science Education
Research, HC Block, Sector III, Salt Lake, Kolkata-700106, India.}

\affiliation{$^{(2)}$ Theoretical Physics Department, Indian
Association for the Cultivation of Sciences, Jadavpur,
Kolkata-700032, India.}

\affiliation{$^{(3)}$ TCMP division, Saha Institute of Nuclear
Physics, 1/AF Bidhannagar, Kolkata-700064, India.}

\begin{abstract}

We obtain the phase diagram of a Bose-Fermi mixture of hardcore
spinless Bosons and spin-polarized Fermions with nearest neighbor
intra-species interaction and on-site inter-species repulsion in an
optical lattice at half-filling using a slave-boson mean-field
theory. We show that such a system can have four possible phases
which are a) supersolid Bosons coexisting with Fermions in the Mott
state, b) Mott state of Bosons coexisting with Fermions in a
metallic or charge-density wave state, c) a metallic Fermionic state
coexisting with superfluid phase of Bosons, and d) Mott insulating
state of Fermions and Bosons. We chart out the phase diagram of the
system and provide analytical expressions for the phase boundaries
within mean-field theory. We demonstrate that the transition between
these phases are generically first order with the exception of that
between the supersolid and the Mott states which is a continuous
quantum phase transition. We also obtain the low-energy collective
excitations of the system in these phases. Finally, we study the
particle-hole excitations in the Mott insulating phase and use it to
determine the dynamical critical exponent $z$ for the
supersolid-Mott insulator transition. We discuss experiments which
can test our theory.

\end{abstract}


\maketitle

\date{\today}

\section {Introduction}

Recent experiments on ultracold trapped atomic gases have opened a
new window onto the phases of quantum matter \cite{Greiner1}. A gas
of Bosonic atoms in an optical or magnetic trap has been reversibly
tuned  between superfluid (SF) and insulating ground states by
varying the strength of a periodic potential produced by standing
optical waves \cite{Greiner1}. This transition has been explained on
the basis of the Bose-Hubbard model with on-site repulsive
interactions and hopping between nearest neighboring sites of the
lattice. \cite{Fisher89,van1}. Further, theoretical studies of
Bosonic atoms with spin and/or pseudospin have also been undertaken
\cite{imambekov1,demler1}. These studies have revealed a variety of
interesting Mott \cite{Greiner1} and supersolid \cite{ssref1} phases
and superfluid-insulator transitions \cite{van1} in these systems.
On the Fermionic side, the experimental studies have mainly
concentrated on the observation of paired superfluid states
\cite{Greiner2} and the BCS-BEC crossover in such systems near a
Feshbach resonance \cite{Zwierlein1}. More recently, it has been
possible to generate mixtures of Fermionic and Bosonic atoms in a
trap \cite{modugno1}. Several theoretical studies followed soon,
which established such Bose-Fermi mixtures  to be interesting
physical systems in their own right \cite{roth1,cramer1}, exhibiting
exciting Mott phases in the presence of an optical lattice.

Many of the earlier studies of Bose-Fermi mixtures has been
restricted to one-dimensional (1D) systems \cite{oned1} or have
concentrated on regimes where the coupling between Bosons and
Fermions are weak \cite{rg1}. The existence of a supersolid (SS)
phase in these system in such a weak coupling regime has also been
predicted \cite{blatter1}. Other works \cite{roth1,cramer1} which
have looked at the strong coupling regime have restricted themselves
to integer filling factors of Bosons and Fermions (per spin) and
have therefore not addressed the phenomenon of translational
symmetry breaking and possible associated SS phases in the strongly
interacting regimes of these systems. More recently, however, the
authors of Ref.\ \onlinecite{hops1} have studied a mixture of
spinless softcore Bosons with an on-site interaction $U$ and
spin-polarized non-interacting Fermions at half-filling in a 3D
optical lattice using dynamical mean-field theory (DMFT). Several
interesting phases, including a SS phase of Bosons and
charge-density wave(CDW) states of Fermions, have been found in
Ref.\ \onlinecite{hops1}.

In this work, we study a mixture of hardcore spinless Bosons and
spin-polarized interacting Fermions in an optical lattice at
half-filling using a slave-boson mean-field technique. We
concentrate on the case where both the Bosons and the Fermions have
a nearest-neighbor density-density repulsive interaction in addition
to the usual on-site interaction term between them. We provide an
analytical, albeit mean-field, phase diagram for the system and
demonstrate that the ground state of such a system consists of four
distinct phases, namely, a) a Mott insulating (MI) phase where both
Fermions and Bosons are localized at the lattice sites, b) a
metal+SF phase where the Fermions are in a metallic phase with a
gapless Fermi surface and the Bosons are in a superfluid state, c) a
SS phase where the Bosons are in the SS state while the Fermions are
localized at the lattice site, and d) a CDW+MI phase of coexisting
Fermions with weak density wave order along with Mott insulating
Bosons. We show, within mean-field theory, that the transition
between these phases are generically first order with the exception
of that between the SS and the MI phases which is a continuous
quantum phase transition. We also obtain the low-energy collective
modes in the metal+SF, CDW+MI and the SS phases and demonstrate that
they have linear dispersions with definite group velocities.
Further, in the MI phase which has no gapless modes, we find the
dispersion of the gapped particle-hole excitations and use it to
determine the dynamical critical exponent $z$ for the continuous
MI-SS transition. We also discuss realistic experiments which can
test our theory.

The plan of rest of the paper is as follows. In Sec.\ \ref{mfa}, we
develop a slave-Boson mean-field theory for the system. This is
followed by Sec.\ \ref{mfb}, where the mean-field phase diagram is
charted out. In Sec.\ \ref{coll}, we obtain the low-energy
collective modes of the system in the metal+SF, CDW+MI and the SS
phases. This is followed by Sec.\ \ref{gm}, where we discuss the
gapped particle-hole excitations of the MI phase and use it to
determine $z$ for the SS-MI transition. We discuss relevant
experiments which can test our theory and conclude in Sec.\
\ref{conc}

\section{Slave-Boson Mean-field Theory}
\label{mft}

\subsection{Mean-field equations}
\label{mfa}

The Hamiltonian of the Bose-Fermi mixture in a $d$-dimensional
hypercubic lattice is described by the Hamiltonian
\begin{eqnarray}
H &=&  H_{F} + H_{B} + H_{FB} \label{sysham} \\
H_F &=& -t_F \sum_{\langle ij\rangle} c_{i}^{\dagger} c_j + V_F \sum_{\langle ij\rangle} n_{i}^F n_j^F
\label{fermiham1} \\
H_B &=& -t_B \sum_{\langle ij\rangle} b_{i}^{\dagger} b_j + V_B \sum_{\langle ij\rangle} n_{i}^B n_j^B
\label{boseham1} \\
H_{FB} &=& U \sum_i n_i^F n_i^B  \label{mixham1}
\end{eqnarray}
where $c_i$ ($b_i$) and $n_i^F=c_i^{\dagger}c_i $ ($n_i^B =
b_i^{\dagger} b_i$) denote the annihilation and number operators for
Fermions(Bosons) at site $i$, $V_F(V_B)$ and $t_F(t_B)$ denote the
nearest-neighbor interaction strengths and hopping amplitudes for
the Fermions(Bosons) respectively, $U$ represents the amplitude for
on-site interaction between the Fermions and the Bosons, and
$\sum_{\langle ij\rangle}$ represents sum over nearest neighbor $ij$
pairs on the lattice. In what follows, we shall study the
Hamiltonian at half-filling. We note at the outset that this
constraint of half-filling implies $n_i^F, n_i^B \le 1$ at each
site.

To obtain an analytical understanding of the phases of these model,
we first introduce a slave-boson representation for the Fermions:
$c_i = a_i d_i$, where $a_i$ denotes annihilation operator for the
slave-bosons and $d_i$ represents the annihilation operator for
pseudo-fermions. We note that in this representation, the
anticommutation relation for the Fermionic operator $[c_i^{\dagger}
c_j ]_+ = \delta_{ij}$, where $\delta_{ij}$ denotes the Kronecker
delta, enforces the constraint $n_{i}^d = n_i^a$ on each site. In
terms of these slave-bosons and pseudo-fermions, the Hamiltonians
$H_F$ and $H_{FB}$ in Eqs.\ \ref{fermiham1} and \ref{mixham1} can be
written as
\begin{eqnarray}
H'_F &=& -t_F \sum_{\langle ij\rangle} d_i^{\dagger} a_i^{\dagger} a_j d_j + V_F
\sum_{\langle ij\rangle} n_i^a n_j^a  \nonumber\\
&& + \sum_i \lambda_i (n_i^a -n_i^d)
\label{fermiham2} \\
H'_{FB} &=& U \sum_i n_i^a n_i^B \label{mixham2}
\end{eqnarray}
where we have implemented the constraint $n_i^a = n_i^d$ using
Lagrange multipliers $\lambda_i$ at each site, and have used the
fact that $n_i^a = n_i^d = n_i^F \le 1$ at each site $i$. We note
that the Hamiltonian $H'= H_B + H'_F + H'_F$ is exact and is
completely equivalent to $H$ (Eq.\ \ref{sysham}).

To make further progress, we proceed with mean-field approximation
of $H'$. To this end, we first decompose the quartic hopping term in
Eq.\ \ref{fermiham2} using
\begin{eqnarray}
t_F \sum_{\langle ij\rangle} d_i^{\dagger} a_i^{\dagger} a_j d_j &=&
t_1 \sum_{ij} a_i^{\dagger}a_j + t_2  \sum_{\langle ij\rangle}
d_i^{\dagger} d_j - \frac{N t_2 t_1}{t_F} \label{mfd1}
\end{eqnarray}
where $N$ is the number of sites in the lattice and the hopping
amplitudes $t_1$ and $t_2$ are given by
\begin{eqnarray}
t_1 &=& \frac{t_F}{N} \sum_{\langle ij \rangle} \langle
d_i^{\dagger} d_j \rangle_{0}, \quad t_2 = \frac{t_F}{N}
\sum_{\langle ij \rangle} \langle a_i^{\dagger} a_j \rangle_{0}.
\label{sc1}
\end{eqnarray}
Here $\langle.. \rangle_0$ denotes average with respect to the
ground state of the system. Next, we use a mean-field approximation
for the constraint term and approximate the Lagrange multiplier
field $\lambda_i = \lambda_0 + \Delta (-1)^i$, where for sake of
definiteness, we take $i=i_1+i_2+...+i_d$ to be even for $A$
sublattice sites. Such an ansatz for $\lambda_i$ is motivated by the
fact that it is the simplest mean-field ansatz that preserves the
basic symmetries of the problem and, at the same time, allows for
translational symmetry breaking in the pseudo-fermion sector. With
these approximations, the mean-field Hamiltonian for the system can
be written as
\begin{eqnarray}
H_{MF} &=& - t_2 \sum_{\langle ij\rangle} d_i^{\dagger} d_j
+ \Delta \sum_i (-1)^i (n_i^a-n_i^d)  \nonumber\\
&& + \sum_{\langle ij\rangle} \left( V_F n_i^a n_j^a -t_1 a_i^{\dagger}  a_j \right)
+ \sum_i U n_i^a n_i^b  \nonumber\\
&& + \sum_{\langle ij\rangle} \left( V_B n_i^B n_j^B -t_B
b_i^{\dagger}  b_j \right) + \frac{N t_2 t_1}{t_F} \label{mfham}
\end{eqnarray}
In writing Eq.\ \ref{mfham}, we have ignored the term $ \lambda_0 \sum_i (n_i^d -n_i^a)$ since
it merely renormalizes the chemical potential of the fermions and thus behave like a constant
as long as we restrict ourselves to half-filling.

To obtain the ground state energy corresponding to this mean-field Hamiltonian, we now use a variational
ansatz for the ground-state wavefunction
\begin{eqnarray}
|\Psi\rangle_0 &=&  \prod_{i \in A} |\psi_A\rangle
\otimes  \prod_{i \in B} |\psi_B\rangle
\otimes |{\rm FS}\rangle \nonumber\\
|\psi_A\rangle &=& \left( \cos(\theta) |n^B=0\rangle + \sin(\theta) |n^B=1\rangle \right)\nonumber\\
&& \otimes  \left( \cos(\gamma) |n^a=0\rangle + \sin(\gamma) |n^a=1\rangle \right) \nonumber\\
|\psi_B\rangle &=& \left( \cos(\theta) |n^B=1\rangle + \sin(\theta) |n^B=0\rangle \right)\nonumber\\
&& \otimes  \left( \cos(\gamma) |n^a=1\rangle + \sin(\gamma) |n^a=0\rangle \right) \nonumber\\
|{\rm FS}\rangle &=& \prod_{\bf k} \theta(|{\bf k}|-k_F) d_k^{\dagger} |0\rangle \label{varwave}
\end{eqnarray}
where $k_F$ denotes the Fermi wavevector for the pseudo-fermions and
$\theta$ and $\gamma$ are variational parameters which has to be
determined by minimizing the ground state energy. Note that the
variational wavefunction given by Eq.\ \ref{varwave} has a two
sublattice structure which allows for the possibility of
translational symmetry broken phases. For the current system, where
the Fermions and the Bosons both interact via nearest-neighbor
density-density interaction terms, Eq.\ \ref{varwave} is the
simplest possible mean-field variational wavefunction which respects
all the symmetries of the Hamiltonian. We would like to point out
here that capturing the phases of a Bose-Fermi mixture with such
interaction terms is beyond the scope of any single-site mean-field
theory including single site DMFT.

The variational mean-field energy $ E_v = \langle \Psi | H_{MF} | \Psi \rangle$ of the system can now be
easily obtained and is given by
\begin{eqnarray}
\frac{E_v}{N t_F} &=& -\frac{d }{2} \left[ Z' \sin^2(2 \gamma) + Z \sin^2(2 \theta) \right] \nonumber\\
&& + \frac{U'}{2} \left[ \cos^2(\gamma) \cos^2(\theta)
+ \sin^2(\gamma) \sin^2(\theta) \right]  \nonumber\\
&& - \frac{\Delta'}{2} \left( \cos(2 \gamma) + \frac{2}{N} \langle
{\rm FS}| \sum_i (-1)^i d_i^{\dagger} d_i |{\rm FS} \rangle \right)
\nonumber\\
Z' &=& (t_1 -V_F)/t_F , \quad Z=(t_B - V_B)/t_F \label{varen}
\end{eqnarray}
where we have used $t_2/t_F =\sin^2 (2 \gamma)$, $\Delta' =
\Delta/t_F$, $U'= U/t_F$, and $t_1/t_F$ has to be determined from
Eq.\ \ref{sc1}. The corresponding mean-field equations which
determine the ground-state values of the variational parameters are
given by $\partial E_v/\partial \theta = \partial E_v/\partial
\gamma = \partial E_v/\partial \Delta' =0$ and yields
\begin{eqnarray}
\sin(2 \theta) \left[ \frac{U'}{2} \cos(2 \gamma) + 2Zd \cos(2 \theta) \right] &=& 0 \nonumber\\
\sin(2 \gamma) \left[ \frac{U'}{2} \cos(2 \theta) + 2Z'd \cos(2 \gamma)
- \Delta' \right] &=& 0 \nonumber\\
\cos(2 \gamma) + \frac{2}{N} \langle {\rm FS}| \sum_i (-1)^i
d_i^{\dagger} d_i |{\rm FS} \rangle &=& 0 \label{mfe1}
\end{eqnarray}

Next, we evaluate the effective hopping amplitude $t_1$ and $ 2/N
\langle {\rm FS}| \sum_i (-1)^i d_i^{\dagger} d_i |{\rm FS}
\rangle$. To this end, first, let us consider the pseudo-fermion
Hamiltonian ${\tilde H} =- t_2 \sum_{\langle ij\rangle}
d_i^{\dagger} d_j - \Delta \sum_i (-1)^i n_i^d$. The energy spectrum
of ${\tilde H}$ is given by $\pm E({\bf k})$, where $E({\bf k})=
\sqrt{ \epsilon({\bf k})^2 + \Delta^2}$, $\epsilon({\bf k}) = -2t_2
\sum_{i=1,d} \cos(k_i a)$ is the energy dispersion of free fermions
in a hypercubic lattice in $d$ dimensions, and $a$ is the lattice
spacing which we shall, from now on, set to unity. The density of
states (DOS) corresponding to these fermions are therefore given by
\begin{eqnarray}
\rho(E) &=& \rho_0(\sqrt{E^2-\Delta^2})
\frac{\sqrt{E^2-\Delta^2}}{E}
\nonumber\\
\rho_0(\tilde \epsilon)  &=& \frac{1}{2t_2} \int \frac{d^d k}{(2
\pi)^d} \delta [ (\tilde \epsilon) -\sum_{i=1}^d \cos(k_i) ]
\label{dos1}
\end{eqnarray}
where $\rho_0$ denotes the DOS of free fermions with tight-binding
dispersion on a hypercubic lattice, $\tilde \epsilon=
\epsilon/2t_2$, and we have used the relation $\rho(E) dE=
\rho_0(\epsilon) d\epsilon$. It is convenient to use Eq.\ \ref{dos1}
to express the expectation values over pseudo-fermion ground states
in Eqs.\ \ref{sc1} and \ref{mfe1} and one obtains
\begin{eqnarray}
\cos(2 \gamma) &=& -{\tilde \Delta} I_1(\tilde \Delta), \quad t_1 =
t_2 I_2 \label{mfe2}\\
I_1(\tilde \Delta) &=& \frac{\int_{-1}^0 \frac{1}{\sqrt{{\tilde
\epsilon}^2 }+ {\tilde \Delta}^2} \rho_0(\tilde \epsilon) d \tilde
\epsilon }{\int_{-1}^0 \rho_0(\tilde \epsilon) d \tilde \epsilon} \nonumber\\
I_2 &=& \frac{\int_{-1}^0 \tilde \epsilon \rho_0(\tilde \epsilon) d
\tilde \epsilon }{\int_{-1}^0 \rho_0(\tilde \epsilon) d \tilde
\epsilon }\label{mfe3}
\end{eqnarray}
Eqs.\ \ref{mfe1}, \ref{mfe2} and \ref{mfe3} denotes the complete set
of mean-field equations which can be now solved to determine the
mean-field phase diagram.

\subsection{Phase diagram}
\label{mfb}

Eqs.\ \ref{mfe1} and \ref{mfe2} can be easily solved numerically to
obtain the mean-field phase diagram for the system. However, before
resorting to numerics, we provide a qualitative discussion of the
nature of the phases.

We find that Eqs.\ \ref{mfe1} and \ref{mfe2} yields four distinct
solutions which corresponds to four possible phases of the system.
First, we find a MI phase with broken translational symmetry where
both the fermions and the Bosons are localized. Such a phase
corresponds to the solution \cite{comment1}
\begin{eqnarray}
\theta &=& 0, \quad \gamma = \pi/2, \quad {\tilde \Delta} \to \infty
\label{sol1}
\end{eqnarray}
Note that the divergence of ${\tilde \Delta}$ corresponds to $t_2
\to 0$ which in turn ensures that $\cos(2\gamma) =-1$. Such a MI
state corresponds to a intertwined checkerboard density-wave pattern
where the Fermions are localized in sublattice $A$ ($n_i^a =
\sin^2(\gamma) = n_i^d = 1$ for $i \in A$) and the Bosons are
localized in sublattice $B$ ($n_i^B=\cos^2(\theta)=1$ for $i \in
B$). The mean-field energy of this state is $E_1=0$.

Second, we find a SS phase, where the Bosons are in a supersolid
phase with coexisting density-wave and superfluid order and the
Fermions are localized in a Mott phase. Such a state corresponds to
the solution
\begin{eqnarray}
\cos(2\theta) &=& \frac{U'}{4Zd}, \quad \gamma=\pi/2,\quad  {\tilde
\Delta} \to \infty \label{sol2}
\end{eqnarray}
Such a state has $\langle b \rangle = \sin(2\theta)/2 \ne 0$ and
$\langle (-1)^i b_i^{\dagger} b_i \rangle = -\cos(2\theta) \ne 0$
and thus corresponds to a SS phase for the Bosons. Note that the
realization of this state necessarily requires $U'/4Zd < 1$. For
$U'/4Zd = 1$, $\theta =0$ and we recover the MI state where $\langle
b\rangle =0$. The energy of the SS state is per site given by
\begin{eqnarray}
E_2 &=& -\frac{Zdt_F}{2}\left(\frac{U'}{4Zd}-1\right)^2 \label{enss}
\end{eqnarray}

Third, we find the MI+CDW state where the fermions show weak
density-wave oscillations whereas the Bosons are localized in the MI
state. This corresponds to the solution
\begin{eqnarray}
\theta &=& 0, \quad \gamma = \gamma_0 \ne 0, \pi/2 \nonumber\\
\end{eqnarray}
where $\gamma_0$ and $\Delta$ are to be determined from a numerical
solution of the mean-field equations
\begin{eqnarray}
\cos(2\gamma_0) &=& -\frac{U'}{4Z'd} \left(1-2\Delta/U\right) =
-{\tilde \Delta} I_1({\tilde \Delta}). \label{mfg}
\end{eqnarray}
The energy of this state per site is given by
\begin{eqnarray}
E_3 = \frac{U\cos^2(\gamma_0)}{2}\left(1-\frac{4Z'd}{U'} \sin^2(\gamma_0)\right)\label{enmicdw}
\end{eqnarray}

Finally, we find the state in which the superfluid Bosons coexist
with metallic Fermions. This corresponds to the solution
\begin{eqnarray}
\theta &=& \gamma=\pi/4 \quad \Delta = 0
\end{eqnarray}
Note that such a phase has  $\langle b \rangle = \sin(2\theta)/2 =1
$ and $\langle (-1)^i b_i^{\dagger} b_i \rangle = -\cos(2\theta) =
0$ so that the Bosons are in an uniform superfluid state. Also,
$\Delta =0$ and $\langle (-1)^i a_i^{\dagger} a_i \rangle =
-\cos(2\gamma) = 0$ in this state indicating that the Fermions are
in a gapless uniform metallic state. The energy of this state per
site is given by
\begin{eqnarray}
E_4 &=& \frac{U}{4} \left[1- \frac{2d}{U'} \left(Z+Z'\right) \right]  \label{enmetal}
\end{eqnarray}

The phase boundaries corresponding to these phases can be
analytically computed using Eqs.\ \ref{enss}, \ref{enmicdw} and
\ref{enmetal}, provided $\gamma_0$ and $t_1$ (which determines $Z'$)
are obtained from numerical solutions of  Eqs.\ \ref{mfg} and Eq.\
\ref{sc1}. For the MI phase to occur, we must have $E_2,E_3,E_4 \ge
E_1=0$ which yields the conditions
\begin{eqnarray}
\left(1- \frac{2d}{U'}(Z+Z')\right) \ge 0, \quad Z\le 0 \, {\rm and} \,
\frac{U'}{4|Z|d} \le 1, \nonumber\\
\frac{4Z'd}{U'} \sin^2(\gamma_0) \le 1 \label{condmott}
\end{eqnarray}
Note that the condition $U'/(4|Z|d)\le 1$ which is necessary for the
realization of the SS phase has to be simultaneously satisfied with
the condition $Z \le 0$ to make sure that the SS phase is actually a
competing candidate to the MI state. The MI phase can indeed be
realized in the parameter regime $Z \ge 0$ provided $U'>4|Z|d$. The
first condition $\left(Z+Z'\right) \le U'/2d$ shows that the MI
phase is favored over the metal+SF phase for large $U/d$ and
predicts a linear phase boundary in the $U'-Z$ plane $U'=2d(Z+Z')$
with a slope of $2d$ and intercept of $2dZ'$ between these two
phases. Note that the MI phase always wins over the metal+SF phase
if the nearest-neighbor interactions between the Bosons and Fermions
are large compared to their hopping amplitudes making $Z+Z'$
negative. The final condition $\frac{4Z'd}{U'} \sin^2(\gamma_0) \le
1$ indicates that the phase boundary between the MI and MI+CDW
phases is independent of $Z$. The former phase is favored over the
latter for larger $U$ and smaller $Z'$.

Similarly for the SS phase to occur one needs $U'/(4|Z|d)< 1$ and
$E_2 \le E_1,E_3,E_4$, which yields
\begin{eqnarray}
\frac{U'}{Z d} \cos^2(\gamma_0) \left(\frac{4Z'd}{U'}
\sin^2(\gamma_0) -1\right)
\le \left(\frac{U'}{4Zd}-1\right)^2 \nonumber\\
Z \ge 0, \quad  \frac{4Z'd}{U'} \sin^2(\gamma_0) \le 1, \quad \frac{U'}{4\sqrt{ZZ'}d} \ge 1
\label{condss} \nonumber\\
\end{eqnarray}
We note that the SS phase is favored when the nearest-neighbor
interaction between the Bosons are weak compared to their hopping
amplitudes making $Z$ positive and when $U'$ is small enough so that
$U'/(4 |Z| d) < 1$. Also, from the conditions in Eq.\ \ref{condss}
(obtained using $E_2 \le E_4,E_1$), we note that for a given $Z'$
and $d$, the boundary between the metal+SF and the SS phases is a
parabola in the $U'-Z$ plane given by $U^{'2}= 16ZZ'd^2$ while that
between the SS and the MI state is a line given by $U'=4Zd$.

Finally, the condition for occurrence of the metal+SF phase is given
by $E_4 \le E_1,E_2, E_3$ and is given by
\begin{eqnarray}
\left(1- \frac{2d}{U'}(Z+Z')\right) \le  0, \quad  \frac{U'}{4\sqrt{ZZ'}d} \ge 1\, {\rm and} \,
\frac{U'}{4|Z|d} \le 1, \nonumber\\
\left(1- \frac{2d}{U'}(Z+Z')\right) \le 2 \cos^2(\gamma_0)
\left(1-\frac{4Z'd}{U'} \sin^2(\gamma_0)\right) \label{condmetal} \nonumber\\
\end{eqnarray}
The last condition in Eq.\ \ref{condmetal} determines the phase
boundary between the metal+SF and the MI+CDW phases which depends on
value of $\gamma_0$. However, numerically, we find that for $U
\simeq 0$ $\gamma_0 \simeq \pi/4$, and in this regime, the phase
boundary between these phases occurs at $Z \simeq0$ for all $Z'$ and
$d$. Note that strictly at $U=0$, the Fermionic state is metallic;
however a CDW gap opens up in the Fermionic spectrum for an
infinitesimal finite $U'$.

\begin{figure}
\centerline{\psfig{file=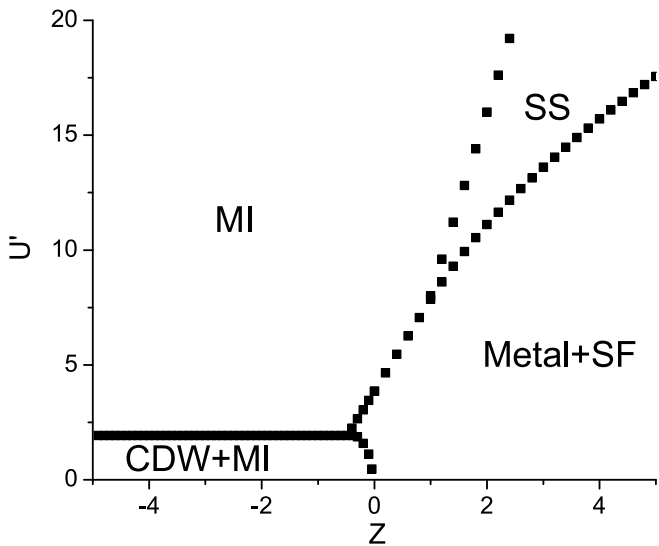,width=\linewidth,angle=0}}
\caption{Ground state phase diagram as a function of $Z$ and $U'$
for noninteracting Fermions ($V_F=0$). The phase boundaries
coincides with the analytical mean-field phase boundaries (see text
for details).} \label{fig1}
\end{figure}

\begin{figure}
\centerline{\psfig{file=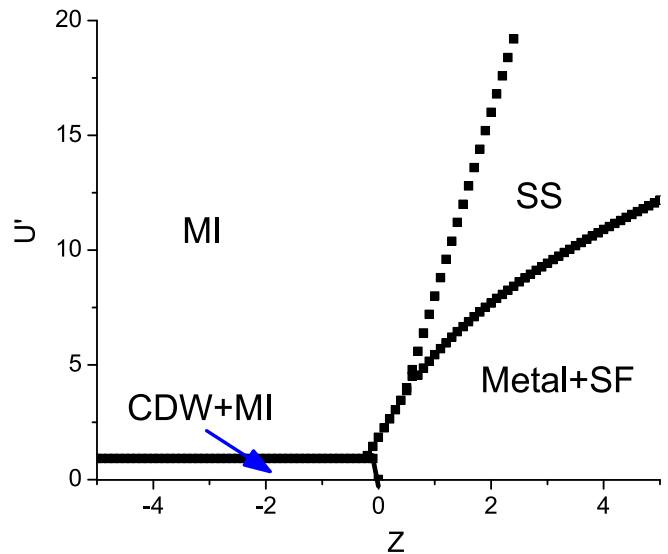,width=\linewidth,angle=0}}
\caption{Same as in Fig.\ \ref{fig1} but for $V_F/t_F=0.5$. The
interaction between the Fermions favors the SS phase as can be seen
by comparing Fig.\ \ref{fig1} and \ref{fig2}.} \label{fig2}
\end{figure}

To verify the above-mentioned qualitative arguments and to find a
precise phase diagram for the system, we numerically solve Eqs.\
\ref{mfe1} ,\ref{mfe2}, and \ref{mfe3}, for $d=2$ and for
representative values $V_F/t_F=0,0.5$. We plot the ground state
phase diagram as a function of $Z$ and $U'$ in Figs.\ \ref{fig1} and
\ref{fig2}. We find that the numerical results agree well with the
qualitative arguments. Figs.\ \ref{fig1} and \ref{fig2} indicate
that the phase boundary between the CDW+MI and MI phases is
independent of $Z$ as noted earlier. The linear and the parabolic
nature of the phase boundaries between the MI and SS phases and the
SS and metal+SF phases respectively can also be easily verified from
the Figs. and are in accordance with the qualitative discussion. We
note that one of the effects of nearest-neighbor repulsion between
the Fermions is to enhance the SS phase which occupies a larger
region of phase space in Fig.\ \ref{fig2} ($V_F/t_F=0.5$) than in
Fig.\ \ref{fig1} ($V_F=0$). Such an interaction, for $Z\le 0$, also
favors the Mott phase over the CDW+MI phase as can also be seen from
Figs. \ref{fig1} and \ref{fig2}.

\begin{figure}
\centerline{\psfig{file=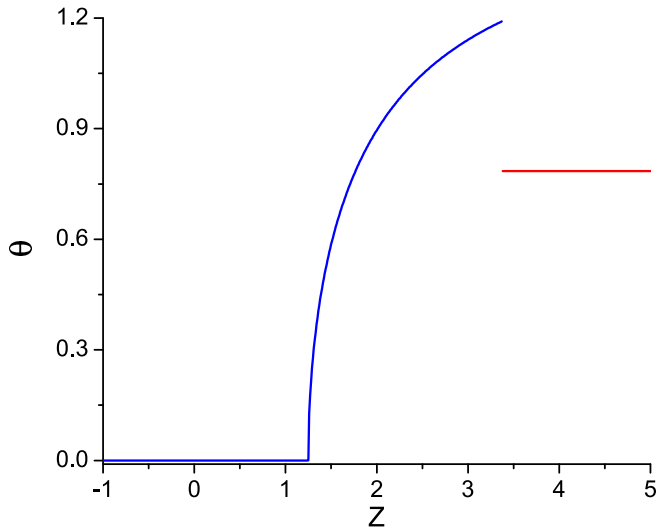,width=\linewidth,angle=0}}
\caption{(Color online) Variation of ground state value of $\theta$
with $Z$ for $U'=10$ and $V_F/t_F=0.5$. The discontinuity in
$\theta$ at $Z=3.38$ indicates a first order transition between the
metal+SF and the SS phases. The transition between the SS and the MI
phases is continuous. } \label{fig3}
\end{figure}

\begin{figure}
\centerline{\psfig{file=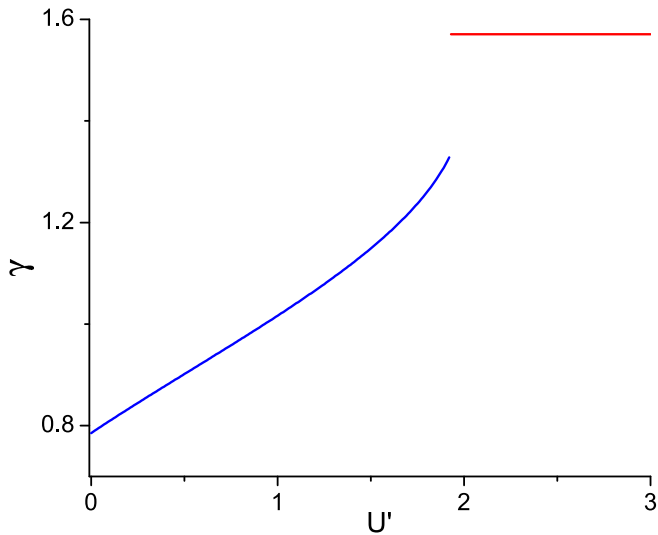,width=\linewidth,angle=0}}
\caption{(Color online) Variation of the ground state value of
$\gamma$ with $U'$ for $Z=-2.$ and $V_F=0$. The discontinuity in
$\gamma$ at $U'=2$ indicates a first order transition between the
CDW+MI and the MI phases.} \label{fig4}
\end{figure}

To determine the nature of transition between the different phases,
we plot the ground state values $\theta$ as a function of $Z$ for
$U'=10$ and $V_F/t_F=0.5$ in Fig.\ \ref{fig3}. Such a plot clearly
shows that the transition between the metal+SF and the SS phases is,
within the mean-field theory considered here, first order and is
accompanied by a jump in the value of $\theta$. In contrast, the
SS-MI transition turns out to be continuous. A similar plot of
ground state values $\gamma$ as a function of $U'$ for $Z/t_F=-2$
and $V_F=0$ , shown in Fig.\ \ref{fig4}, indicates that the
transition between the CDW+MI and the MI phase is also discontinuous
and is accompanied by a jump in the ground state value of $\gamma$.

%%%%%%%%%%%%%%%%%%%%%%%%%%%%%%%%%%%%%%%%%%%%%%%%%%%%%%%%%%%
\noindent
%%%%%%%%%%%%%%%%%%%%%%%%%%%%%%%%%%%%%%%%%%%%%%%%%%%%%%%%%%%
%%%%%%%%%%%%%%%%%%%%%%%%%%%%%%%%%%%%%%%%%%%%%%%%%%%%%%%%%%%
\noindent
%%%%%%%%%%%%%%%%%%%%%%%%%%%%%%%%%%%%%%%%%%%%%%%%%%%%%%%%%%%

Next, we compare our phase diagram with that obtained from DMFT in
Ref.\ \onlinecite{hops1}. This can be done in the regime of large
positive $Z$ (which correspond to $V_B/t_B \to 0$) which was the
case treated in Ref.\ \onlinecite{hops1}. We find that the two phase
diagrams qualitatively agree in the sense that both yield SS and MI
phases in these limit. The difference lies in the fact that our
mean-field predicts a second-order transition between the two phases
whereas DMFT yields a narrow region of coexistence. This is
presumably an effect of quantum fluctuation which is not captured
within the mean-field theory. In addition, we also find a region of
metal+SF phase at low $U$ which was not seen in Ref.\
\onlinecite{hops1}.

Finally, we would like to point out that the slave-boson mean-field
phase diagram obtained above yields qualitatively correct phase
diagram, but not a quantitatively correct one. This can be most
clearly seen by noting that our Hamiltonian reduces to an effective
Falicov-Kimball (FK) model \cite{fk1} in the limit
$Z=t_B=V_B=V_F=0$. This is most easily seen by writing our starting
Hamiltonian $H$ (Eq.\ \ref{sysham}) for $t_B=V_B==V_F=0$
\begin{eqnarray}
H_{FK} &=& -t_F \sum_{\langle ij \rangle} c_i^{\dagger} c_j + U
\sum_i c_i^{\dagger} c_i n_i^B  \label{fk1}
\end{eqnarray}
At half-filling the Bosons are localized in the B sublattice so that
$n_i^B =(1-(-1)^i)/2$. Thus the Fermions have a CDW instability even
for an infinitesimal $U$ due to nesting for half-filling on a square
lattice. Consequently, the FK model at half-filling is insulating
for any infinitesimal $U$, as known from several earlier studies
\cite{fk2}. Such a CDW instability, which can be easily captured by
weak-coupling mean-field theory, is not straightforward to obtain in
our strong coupling slave-boson mean-field approach which predicts a
finite critical $U$ for the transition from metal+SF to the MI
phase. Note however that the slave-boson mean-field theory does
predict a CDW+MI state for weak $U$, but for small negative $Z$, as
can be seen from Fig.\ \ref{fig1}. This indicates that the phase
diagram obtained has qualitatively, but not quantitatively, correct
features.

%%%%%%%%%%%%%%%%%%%%%%%%%%%%%%%%%%%%%%%%%%%%%%%%%%%%%%%%%%%
\noindent
%%%%%%%%%%%%%%%%%%%%%%%%%%%%%%%%%%%%%%%%%%%%%%%%%%%%%%%%%%%
%%%%%%%%%%%%%%%%%%%%%%%%%%%%%%%%%%%%%%%%%%%%%%%%%%%%%%%%%%%
\noindent
%%%%%%%%%%%%%%%%%%%%%%%%%%%%%%%%%%%%%%%%%%%%%%%%%%%%%%%%%%%

\section{Collective Modes}
\label{coll}

The phases of the Bose-Fermi mixture discussed in the previous
section allows for two types of excitations. The first type is the
low-energy gapless collective modes that are present in the
metal+SF, CDW+MI, and the SS phases of the system. These are the
gapless Goldstone modes corresponding to the Boson and the
slave-Boson fields. For all of these modes, the pseudo-Fermion
sector remain gapped and do not contribute to their dispersion. The
gapped modes corresponds to particle-hole excitations in the CDW+MI,
SS and the Mott states. In this section, we concentrate on the
gapless collective modes in the CDW+MI, SS, and the metal+SF phases.

To obtain the dispersion, we first consider a time-dependent
variational wave-function
\begin{eqnarray}
|\Psi_d(t)\rangle_0 &=&  \left( \cos(\theta_i) |n_i^B=0\rangle + \sin(\theta_i)
e^{-i\chi_i}|n_i^B=1\rangle \right)\nonumber\\
&& \otimes  \left( \cos(\gamma_i) |n_i^F=0\rangle + \sin(\gamma_i) e^{-i\phi_i} |n_i^F=1\rangle \right)
\label{varwavecol} \nonumber\\
\end{eqnarray}
where $\theta_i$, $\chi_i$, $\gamma_i$, and $\phi_i$ are space-time
dependent fields. Note that in the static limit, the ground-state of
the system corresponds to
$\theta_i=\theta_{i0}=\theta_0(\pi/2-\theta_0)$,
$\gamma_i=\gamma_{i0}= \gamma_0(\pi/2-\gamma_0)$ for $i \in A(B)$
sites, and $\chi_i=\phi_i=0$, so that $|\Psi_d(t)\rangle$ reduces to
$|\Psi \rangle$.

The Lagrangian ${\mathcal L} = \sum_i \langle \Psi_d(t)| i
\partial_t - H' +\mu_B \sum_i n_i^B +\mu_F \sum_i n_{I}^F| \Psi_d
\rangle$ can now be computed using the variational wave-function and
one obtains
\begin{eqnarray}
{\mathcal L} &=& \sum_{i} \Big[ \partial_t {\chi}_i \sin^2(\theta_i)
+ \partial_t {\phi}_i \sin^2(\gamma_i) \nonumber\\
&& - U \sin^2(\theta_i)\sin^2(\gamma_i) \Big] \nonumber\\
&& + \sum_{\langle ij\rangle} \Big[ \frac{1}{4} \left( t_B \sin(2\theta_i) \sin(2 \theta_j)
\cos(\chi_i-\chi_j)\right. \nonumber\\
&& \left. + t_1 \sin(2\gamma_i) \sin(2 \gamma_j) \cos(\phi_i-\phi_j)  \right. \nonumber\\
&& \left. - \{ V_F \sin^2(\gamma_i) \sin^2(\gamma_j)
+ V_B \sin^2(\theta_i) \sin^2(\theta_j) \} \right) \Big] \nonumber\\
&& + \sum_{i} \Big[ \Delta \left\{- \sin^2(\gamma_i)  (-1)^i +
\langle n_i^F\rangle \right\}  \nonumber\\
&& +\mu_B \sin^2(\theta_i) +\mu_F \sin^2(\gamma_i) \Big]
\end{eqnarray}
where $\langle n_i^F \rangle=\langle FS| (-1)^i d_{i}^{\dagger} d_i
|FS \rangle/N$ is the Fermion number density different on $A$ and
$B$ sublattices and all time dependence of the fields are kept
implicit for the sake of clarity. To obtain the collective modes, we
now write $\theta_i(t) = \theta_{i0} + \delta \theta_i (t)$,
$\gamma_i (t) = \gamma_{i0} + \delta \gamma_i (t)$, and expand the
Lagrangian to quadratic order in $\delta \theta_i(t)$, $\delta
\gamma_i(t)$, $\phi_i(t)$ and $\chi_i(t)$. Then a variation of this
Lagrangian with respect to $\delta \theta_i(t)$, $\delta
\gamma_i(t)$, $\phi_i(t)$ and $\chi_i(t)$ and consequent adjustment
of values of the parameters $\mu_B$ and $\mu_F$ following Ref.\
\onlinecite{sinha1}, yields the equations for the low-energy
collective modes (we set $\hbar=1$ from now on)
\begin{eqnarray}
\partial_t {\delta \gamma}_{\bf k} + t_1 d \sin(2\gamma_0) \left(1-c({\bf k})\right) \phi_{\bf k}
&=& 0 \label{geq1}  \\
\partial_t {\phi}_{\bf k} - \alpha_1({\bf k}) \delta \gamma_{\bf k} - U \sin(2\theta_0) \delta
\theta_{\bf k} &=& 0 \label{geq2} \\
\partial_t {\delta \theta}_{\bf k} + t_B d \sin(2\theta_0) \left(1-c({\bf k})\right) \chi_{\bf k}
&=& 0 \label{geq3}  \\
\partial_t {\chi}_{\bf k} - \alpha_2({\bf k}) \delta \theta_{\bf k} - U \sin(2\gamma_0) \delta \gamma_{\bf k}
&=& 0 \label{geq4}
\end{eqnarray}
where we have taken the Fourier transform of all the fields, $c({\bf
k}) = \sum_{j=1,d} \cos(k_j)/d$, and $\alpha_1$ and $\alpha_2$ are
given by
\begin{eqnarray}
\alpha_1({\bf k}) &=& \frac{4t_1 d}{\sin(2\gamma_0)} \Big [1+ c({\bf k})
\Big(\frac{V_F}{t_1} \sin^2(2\gamma_0)  \nonumber\\
&& + \cos^2(2\gamma_0) \Big) \Big] \nonumber\\
\alpha_2({\bf k}) &=& \frac{4t_B d}{\sin(2\theta_0)} \Big[1+ c({\bf k})
\Big (\frac{V_B}{t_B} \sin^2(2\theta_0) \nonumber\\
&& + \cos^2(2\theta_0) \Big) \Big]  \label{cons1}
\end{eqnarray}
It is important to note that Eqs.\ \ref{geq1} and \ref{geq2} holds
when $\gamma_0 \ne 0,\pi/2$ while Eqs.\ \ref{geq3} and \ref{geq4}
holds when $\theta_0 \ne 0,\pi/2$. Thus none of these equations are
valid in the MI phase which do not support any low-energy collective
modes. The gapped modes of the MI phases will be obtained in the
next section.

In the SS phase where $\cos(2 \theta_0) = U'/(4Zd) \ne 0,1$ and
$\gamma_0 = \pi/2$, the collective mode corresponds to the
low-energy excitations of the Bosons and are given by Eqs.\
\ref{geq3} and \ref{geq4}. A simple set of standard manipulations of
these equations yield the dispersion of the collective modes
$\omega^2 = 2 v_1^2({\bf k}) (1-c({\bf k}))$, where
\begin{eqnarray}
v_1^2({\bf k}) &=& t_B d \sin(2 \theta_0) \alpha_2({\bf k})/2
\label{vel1}
\end{eqnarray}
Note that for low momentum, we get a gapless linearly dispersing
collective mode with velocity $v_{ss}= v_1({\bf k=0})$. Similarly
for the MI+CDW phase, where $\theta_0=0$ and $\gamma = \gamma_0$,
the collective mode corresponds to the low energy excitations of the
pseudo-bosons and can be obtained by solving Eqs.\ \ref{geq1} and
\ref{geq2}. Since the pseudo-Fermion sector is always gapped in this
phase ($\Delta \ne 0$), the collective mode here corresponds to the
density-wave mode of the real Fermions. These modes have linear
dispersion $\omega^2  = 2 v_2^2({\bf k}) (1 - c({\bf k}))$ where
\begin{eqnarray}
v_2^2({\bf k}) =  t_1 d \sin(2 \gamma_0) \alpha_1({\bf k})/2
\label{vel2}
\end{eqnarray}
Thus for low momenta, we again get a gapless linearly dispersing
collective mode with velocity $v_{CDW}= v_2({\bf k=0})$.

Finally for the metal+SS phase, all the Eqs.\ \ref{geq1}..\ref{geq4}
hold and they need to be solved simultaneously. In this phase since
$\gamma_0= \theta_0 = \pi/4$, we find that $\alpha_1({\bf k})= 4t_1
d [1+ c({\bf k})V_F/t_1]$ and $ \alpha_2({\bf k})= 4t_B d [1+ c({\bf
k})V_B/t_B]$. Solving these equations, one finds two collective
modes with linear dispersions $\omega_{\pm}^2 = 2 v_{\pm}^2 ({\bf k}
(1-c({\bf k})) $ where $v_{\pm} ({\bf k})$ are given by
\begin{eqnarray}
v_{\pm}^2({\bf k}) =  \frac{1}{4} \Big[ (\alpha_1({\bf k})t_1 d  + \alpha_2({\bf k})t_B d)
\nonumber\\
\pm \sqrt{ (\alpha_1({\bf k})t_1 d - \alpha_2({\bf k}) t_B d)^2 + 16(Ut_B t_1 d)^2} \Big]
\label{vel3}
\end{eqnarray}
These collective modes result from the hybridization of the
Bogoliubov modes of the Bosons and the density-wave modes for the
metallic Fermions. This fact can be easily checked by putting $U=0$
in Eq.\ \ref{vel3} by which one can retrieve these modes with
velocities $v_{B}^2( {\bf k}) = \alpha_2({\bf k}) t_B d/2$ for the
Bosons and $v_F^2 ({\bf k})= \alpha_1({\bf k}) t_1 d/2$ for the
Fermions. As $U'$ increases, the hybridization between these modes
become stronger until the velocity $v_{-}({\bf k}= \pi)$ touches
zero at $U'= 4 \sqrt{|Z'Z|} d$ which is precisely the condition for
the metal+SF phase to become unstable to the SS phase.

\section{Gapped Modes in the MI phase}
\label{gm}

\begin{figure}
\centerline{\psfig{file=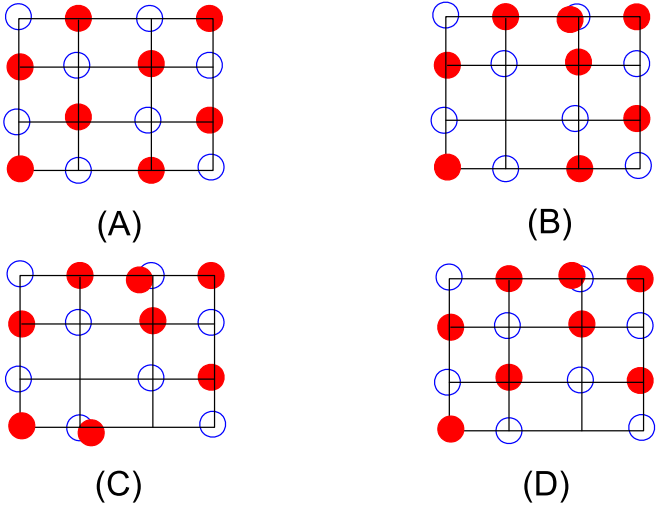,width=\linewidth,angle=0}}
\caption{(Color online) Cartoon representation of the MI state. A)
The Mott state at half-filling. The red filled circles represent
Bosons and the empty blue circles indicate Fermions. B) A particle
and a hole excitation which are not nearest neighbors. C) An
intermediate virtual high energy state which assists hopping of
holes. D) A state where the hole has hopped to the next-neighbor
site. This state is identical in energy to the state B. }
\label{fig5}
\end{figure}

\begin{figure}
\centerline{\psfig{file=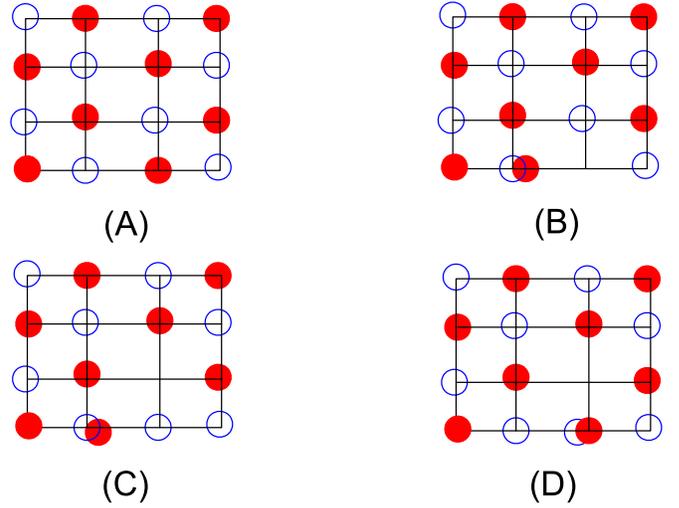,width=\linewidth,angle=0}}
\caption{(Color online) Cartoon representation of the MI state and
the associated dipole excitations. All symbols are same as in Fig.\
\ref{fig5}. A) The Mott state at half-filling. B) A dipole
excitation over the Mott state C) An intermediate virtual high
energy state which assists hopping of dipoles. D) A state where the
dipole has hopped to an adjacent link. This state is identical in
energy to the state B when $V_F=V_B$.} \label{fig6}
\end{figure}

The MI phase, in contrast to the other three phases of the system,
do not support a gapless mode. The lowest-lying excitations of such
a state with conserved number density are particle-hole excitations.
Such excitations, are of two types. The first type, shown in second
panel of Fig.\ \ref{fig5}, involves particle and hole excitations
that spatially well-separated while the second type, shown in second
panel of Fig.\ \ref{fig6}, involves particle and hole excitations in
nearest-neighbor sites which forms a dipole. In what follows, we
first compute the energies of both these excitations using
perturbation theory up to second order in $t_{B/F}/V_{B/F}$ which
are supposed to small in the MI phase.

Such an energy estimate can be easily carried out by strong-coupling
perturbation theory developed in Ref.\ \onlinecite{freemon} in
context single species Bose-Hubbard model. The generalization is
largely trivial, except for one important detail. In the standard
Bose-Hubbard model studied in Ref.\ \onlinecite{freemon}, any
particle/hole excitation could have lowering of energy via
nearest-neighbor hopping which is ${\rm O}(t/U)$ process. In
contrast, as can be seen from Figs.\ \ref{fig5} and \ref{fig6}, it
is not possible for the particle-hole or dipole excitations to
directly hop to the next site since such a direct hop always take us
out the low energy manifold of states in the Mott limit. In
particular, we note that any kinetic energy gain of the
particle-hole or dipole excitation must occur via hopping of the
partice/hole to the second-neighbor sites and hence necessarily
leads to ${\rm O}(t^2/V^2)$ energy gain.
%%%%%%%%%%%%%%%%%%%%%%%%%%%%%%%%%%%%%%%%%%%%%%%%%%%%%%%%%%%
\noindent
%%%%%%%%%%%%%%%%%%%%%%%%%%%%%%%%%%%%%%%%%%%%%%%%%%%%%%%%%%%

We first compute the excitation energy of the Bosonic(Fermionic)
particle-hole pair when they are far apart. The on-site energy of
creating such a pair is $E_{{\rm on-site}}^{B/F} = 4d V_{B/F} +U$
while the energy-lowering due to hopping of each of the particle and
the hole is given by $E_{{\rm hopping}}^{B/F} =
-2d(2d-1)t_{B/F}^2/[2(2d-2)V_{B/F}+U]$. The energy of the Mott state
to second order in perturbation theory is $E_{MI} =
-d(t_B^2/[2(2d-1)V_B+U] + t_F^2/[2(2d-1)V_F+U])$ so that the
excitation energy of the particle-hole pair is
\begin{eqnarray}
E^{\rm B/F}_{p-h} &=&  4d V_{B/F} +U
-\frac{4d(2d-1)t_{B/F}^2}{2(2d-2)V_{B/F}+U}
\nonumber\\
&& + \frac{d t_B^2}{2(2d-1)V_B+U} + \frac{d t_F^2}{2(2d-1)V_F+U}
\label{enph}
\end{eqnarray}
We note that in the limit of large $d$, where the mean-field results
are expected to be accurate, we have
\begin{eqnarray}
E^{\rm B/F}_{p-h} \simeq 4d V_{B/F} +U -\frac{8d^2 t_{B/F}^2}{4d
V_{B/F}+U} \label{enphh}
\end{eqnarray}
The Mott state is destabilized in favor of the SS phase when $E^{\rm
B/F}_{p-h}=0$.
%%%%%%%%%%%%%%%%%%%%%%%%%%%%%%%%%%%%%%%%%%%%%%%%%%%%%%%%%%%
\noindent
%%%%%%%%%%%%%%%%%%%%%%%%%%%%%%%%%%%%%%%%%%%%%%%%%%%%%%%%%%%

Next, we compute the excitation energy of the Bosonic/Fermionic
dipole state. We are going to do this in the limit of $V_F=V_B=V$.
We note at the outset that once such a dipolar excitation is
created, it remains stable, {\it i.e.}, the hole can not hop away
arbitrarily far away from the particle. It can be easily verified
from Fig.\ \ref{fig6} that such hoppings generate higher-energy end
states and takes one out of the low energy manifold of states.

The on-site energy cost for creating such an excitation is
$E_{on-site}= 2(2d-1)V+U$ while the hopping process, shown
schematically in Fig.\ \ref{fig6}, necessarily involves hopping of
both Fermions and Bosons and leads to an energy gain of $E_{{\rm
hopping}}^d= t_B t_F/2(2d-1)V$. Thus the net energy of such a dipole
excitation is given by
\begin{eqnarray}
E_{dipole} &=&  2(2d-1)V+U -\frac{t_{B}t_F}{2(2d-1)V}
\nonumber\\
&& + \frac{d t_B^2}{2(2d-1)V+U} + \frac{d t_F^2}{2(2d-1) V+U}
\label{endi}
\end{eqnarray}
We note that for $d \gg 1$, where our mean-field theory holds, it is
always energetically favorable to create particles and holes
well-separated since $E_{{\rm hopping}}^{B/F} \ll E_{{\rm
hopping}}^d$. However, the dipolar excitations may becomes favorable
in low dimension and for large $U/V$. In this case, $E_{{\rm
hopping}}^{B/F} \gg E_{{\rm hopping}}^d$ for $U \gg V$ and in this
limit, the dipolar excitations would be preferred in destabilizing
the MI phase. We shall not discuss this issue here any further since
this is clearly beyond the scope of our mean-field theory.

Finally, we compute the dispersion of the gapped particle-hole
excitations within mean-field theory where the particle and the hole
are spatially well-separated and do not interact. To this end, we
temporarily relax the constraint of conservation of particle-number
and consider the energy of excitations of adding a particle $E_p$
and a hole $E_h$ to the Mott state. The physical particle-hole
excitation energy can then be computed from $E_{ph}=E_p +E_h$. To
compute the energy of these particel/hole excitations, we adapt a
time-dependent variational approach as done in Ref.\
\onlinecite{sinha1} for single species Bosons in an optical lattice.
We begin with the variational wave-function of Bosons and slave
Bosons(Fermions) given by
\begin{eqnarray}
|\psi(t)\rangle &=& |\psi_B(t)\rangle \times |\psi_F(t)\rangle \nonumber\\
|\psi_B(t)\rangle &=&  f_{0}^{\alpha}(t)|n^B=0\rangle + f_{1}^{\alpha}(t)|n^B=1\rangle \nonumber\\
|\psi_F(t)\rangle &=& g_{0}^{\alpha}(t)|n^a=0\rangle + g_{1}^{\alpha}(t)|n^a=1\rangle \label{wave1}
\end{eqnarray}
where $\alpha=A,B$ denotes sublattice indices. The coefficients $f$
and $g$ satisfy the normalization condition $|f_0|^{2} + |f_1|^{2} =
1$ and $ |g_0|^{2} + |g_1|^{2} = 1$. We note that at equilibrium
$f_{0}^{A} = \sin\theta =0$, $f_{1}^{A} = \cos\theta=1$, $g_{0}^{A}
= \cos\gamma=0$, $g_{1}^{A} = \sin\gamma=1$, $f_{0(1)}^{B}=
f_{1(0)}^A$, and  $g_{0(1)}^{B}= g_{1(0)}^A$ for the MI phase.

The Lagrangian of the Bose-Fermi mixture can then be written as
\begin{eqnarray}
L' &=& \sum_j  i \left[f_{0j}^{*} \dot{f}_{0j} +
f_{1j}^{*}\dot{f}_{1j} +g_{0j}^{*}\dot{g}_{0j}
+ g_{1j}^{*}\dot{g}_{1j}\right] \nonumber\\
&& - \sum_{\langle ij \rangle} \Big[-t_B  f_{1i}^{*}f_{0i}
f_{0j}^{*}f_{1j}
-t_{1} g_{1i}^{*}g_{0i} g_{0j}^{*}g_{1j} \nonumber\\
&& + V_B |f_{1i}|^{2} |f_{1j}|^{2} + V_F  |g_{1i}|^{2}
|g_{1j} |^{2} \Big] \nonumber\\
&& - U \sum_i |f_{1i}|^{2} |g_{1i}|^{2} + \mu_{b} \sum_i
|f_{1i}|^{2}  \nonumber\\
&& +\mu_{f} \sum_i |g_{1i}|^{2} -\sum_i \lambda_{i} (|f_{0i}|^{2} + |f_{1i}|^{2} -1) \nonumber\\
&&  -\sum_i \nu_{i} (|g_{i0}|^{2} + |g_{1i}|^{2}-1) \label{mila1}
\end{eqnarray}
where $\lambda_i$ and $\nu_i$ are variational parameters used for
implementing the constraint whose values are to be determined from
proper choice of the saddle point which in the MI phase yields
$\lambda_{A} = \mu_{b}$ and $\lambda_{B} = 0$ for the Bosons and
$\nu_{A} = 0$ and $\nu_{B} = \mu_{f}$ for the slave bosons.

The saddle-point equations for the variational coefficients $f_{i}(t)$ and $g_{i}(t)$
then reads
\begin{eqnarray}
i\dot{f}_{0i} +t_B f_{1i} \sum_{\langle j\rangle} f_{1j}^{*} f_{0j}
- \lambda_{i} f_{0i} &=& 0
\nonumber\\
i\dot{g}_{0i} +t_1 g_{1i} \sum_{\langle j\rangle} g_{1j}^{*} g_{0j}
- \nu_{i} g_{0i} &=& 0
\nonumber\\
i\dot{f}_{1i} - \Big[ -t_B f_{0i} \sum_{\langle j\rangle} f_{0j}^{*}
f_{1j} + 2V_B f_{1i} \sum_{\langle j\rangle}
|f_{1j}|^{2}  \nonumber\\
+ U f_{1i} |g_{1i}|^{2} - \mu_{b} f_{1i} \Big]
- \lambda_{i} f_{1i} =  0 \nonumber\\
i\dot{g}_{1i} - \Big[ -t_1 f_{0i} \sum_{\langle j\rangle} g_{0j}^{*}
g_{1j} + 2V_F g_{1i} \sum_{\langle j\rangle}
|g_{1j}|^{2}  \nonumber\\
+ U g_{1i} |f_{1i}|^{2} - \mu_{f} g_{1i} \Big] - \nu_{i} g_{1i} = 0
\label{fequ}
\end{eqnarray}

Next we implement the two sublattice structure, shift to momentum
and frequency space, and expand $f_{a{\bf k}}^{A/B}= \delta f_{a
{\bf k}}^{A/B} + f_{a}^{A/B}$ and $g_{a{\bf k}}^{A/B}= \delta g_{a
{\bf k}}^{A/B} + g_{a}^{A/B}$ where $a=0,1$. Note that since $\delta
f^{A/B}$ and $\delta g^{A/B}$ corresponds to deviation of particle
number of the MI state, these dispersion corresponding to their
eigenmodes must represent the particle and hole excitations over the
MI phase. In the MI phase, we find that the equation of motions for
the Bosons and the pseudobosons decouple at linear order  $\delta
f_{a {\bf k}}^{A/B}$ and $\delta g_{a {\bf k}}^{A/B}$. For the
Bosons, we obtain, to linear order in $\delta f_{a {\bf k}}^{A/B}$
\begin{eqnarray}
-\omega \delta f_{0{\bf k}}^{A} &=& -2 d t_B c({\bf k}) \delta f_{1{\bf k}}^{*B}
+ \lambda_{A} \delta f_{0{\bf k}}^{A}\nonumber\\
-\omega \delta f_{1k}^{B} &=& -2 t_B d c({\bf k}) \delta f_{0{\bf k}}^{A} +
2V_{B}z \delta f_{1{\bf k}}^{B} \nonumber\\
&& + U \delta f_{1{\bf k}}^{B} -(\mu_{b} - \lambda_{B}) \delta
f_{1k}^{B}
\end{eqnarray}
which yields two physical excitation dispersion corresponding to particle and
hole excitations
\begin{eqnarray}
E_{p(h)}&=& +(-)(2 d V_{B} + \frac{U}{2} - \mu_{b}) \nonumber\\
&& + \sqrt{(2d V_{B} +
\frac{U}{2})^{2} - (2t_B d c({\bf k}))^2 } \label{phex1}
\end{eqnarray}
The energy of a particle-hole excitation which conserves particle number is therefore
obtained by adding $E_p$ and $E_h$ and is given by
\begin{eqnarray}
E_{p-h} = 2 \sqrt{(2d V_{B} + \frac{U}{2})^{2} - (2t_B d c({\bf k}))^2}
\label{phex2}
\end{eqnarray}
Note that $E_{p-h}$ vanishes along the line $Z=U'/4d$ which agrees
with the mean-field result for the SS-MI phase boundary. Also,
expanding Eq.\ \ref{phex2} to ${\rm O} (t_B^2)$ for ${\bf k}=0$
leads exactly to Eq.\ \ref{enphh} which shows that the second-order
perturbation theory discussed earlier agrees to the present
calculation in the high $d$ limit. Further, at small wave-vector, we
find $E_{p-h} \sim |{\bf k}|$ which shows that the SS-MI quantum
phase transition has a dynamical critical exponent $z=1$. Similar
dispersion can be obtained for the pseudo-bosons by considering
collective modes corresponding to $\delta g_{a{\bf k}}$. These modes
have the same dispersion as Eqs.\ \ref{phex1} and \ref{phex2} with
$V_B$ and $t_B$ replaced by $V_F$ and $t_1$ respectively.

\section{Discussion}
\label{conc}

Experimental realization of Bose-Fermi mixtures have long been
achieved in ultracold atomic systems. These mixtures can be easily
tuned to a regime where the on-site intra-species interaction
between both the Fermions and the Bosons are large so that they
effectively behave as hard-core particles. Thus experimental
realization of a Bose-Fermi mixture with $V_B=V_F=0$ is relatively
straightforward. However, most such mixtures do not have
sufficiently large nearest-neighbor repulsion and thus it might be
difficult to realize mixtures which has large $V_F$ or $V_B$. Some
progress in this direction has recently been made in Ref.\
\onlinecite{exp0}. Also, use of spin-polarized $^{52}{\rm Cr}$ atoms
for the Fermionic part of the mixture may help since these atoms
have significant dipole moment which may provide the requisite
interaction.

Once such a Bose-Fermi system is realized, several predictions of
the present work can be verified by realizable experiments that are
commonly used for ultracold systems. First, we note that since the
Bosons are spinless and the Fermions are spin-polarized, the Bosonic
and the Fermionic part of the mixture can be separated by applying a
standard Stern-Gerlach field during a standard time-of-flight
experiment as done earlier in Ref.\ \onlinecite{exp1} in the context
of spinor Bosons in optical traps. Such a procedure allows us to
separately study the momentum distribution functions of the Bosons
and the Fermions using time-of-flight experiments \cite{Greiner1}.
For the Bosonic cloud, the distinction between the SF and the MI
phases can be easily done by measuring the presence or absence of
coherence peaks in its momentum distribution as measured in a
standard time-of-flight experiment. The precise nature of the broken
translational symmetry in the MI and the SS phases for the Bosons
can also be determined by studying noise-correlations of the
expanding clouds as already proposed in Ref.\ \onlinecite{exp2}.
Thus, the MI, SS and the SF phases for the Bosons can be
qualitatively distinguished by these experiments. As for the
Fermions, the presence/absence of a Fermi surface for the Fermions
in a trap can be easily distinguished in time-of-flight measurements
as performed for ultracold Fermions in Ref.\ \onlinecite{exp3}.
Thus, these experiments should allow one to qualitatively
distinguish between all four predicted phases. One of the central
predictions of our theory is that for any finite $U$, the metallic
state of the Fermions shall always be accompanied by a SF phase of
the Bosons. In terms of time-of-flight experiments this means that
any measurement on Fermions which sees a gapless Fermi surface shall
always be accompanied by corresponding coherence peak (and no
density wave ordering) for the Bosons. The collective modes computed
in this work can also be verified experimentally using standard
inelastic light scattering experiments \cite{exp4}. Such experiments
can detect low-energy collective modes and should thus detect either
two ( metal+SF phase) or one (SS or MI+CDW phases) linearly
dispersing collective mode(s). The MI phase will be characterized by
absence of any low energy collective modes of the system.

There are several possible extension of our analysis. The first and
the simplest extension would be to study the phases of the
Bose-Fermi mixture away from half-filling. This would require a more
careful handling of the chemical potential $\mu_B$ and $\mu_F$ of
the Bosons and the Fermions. In particular one would need to
determine $t_1$ in a self-consistent manner as a function of
$\mu_F$. Second, it would be interesting to look at the phase
diagram by relaxing the hard-core constraint on the Bosons by
putting a finite on-site repulsion between them. Of particular
interest in this respect is to check if the slave-boson mean-field
can provide any indication of the phase separation found in such a
system in Refs.\ \onlinecite{blatter1,hops1}. Finally, it would be
useful to study the phase diagram of mixture of spin-polarized
Fermions with spin-one and spin-two Bosons with nearest-neighbor
interactions. Such system are expected to have a much richer phase
diagram and have not been theoretically studied so far.

To conclude, in this work, we have carried out a slave-boson
mean-field analysis of a mixture of hardcore spinless Bosons and
spin-polarized Fermions in an optical lattice. Our analysis provides
the mean-field phase diagram of the system and shows the presence of
four distinct phases. We have also computed the low-energy
collective modes of three of these phases (metal+SF, CDW+MI and SS)
and studied the gapped particle-hole excitation of the fourth (MI).
We have discussed experiments which can be used to test our theory
and possible extension of our theory to other systems.

We thank Jim Freericks for drawing our attention to the
Falicov-Kimball limit of the present model and for several useful
discussions.


\begin{thebibliography}{99}
\bibitem{comment1} Alternatively, we could have chosen the solution
$\theta=\pi/2$, $\gamma=0$ and $\Delta \to -\infty$. This simply
amounts to relabeling of A and B sublattices and corresponds to the
same physical state.
\end{thebibliography}
\end{document}